\documentclass[twocolumn]{aastex631}

\shorttitle{Eccentricity Driven Climate Effects}
\shortauthors{Stephen R. Kane et al.}

\begin{document}

\title{Eccentricity Driven Climate Effects in the Kepler-1649 System}

\author[0000-0002-7084-0529]{Stephen R. Kane}
\affiliation{Department of Earth and Planetary Sciences, University of
  California, Riverside, CA 92521, USA}
\email{skane@ucr.edu}

\author[0000-0002-4860-7667]{Zhexing Li}
\affiliation{Department of Earth and Planetary Sciences, University of
  California, Riverside, CA 92521, USA}

\author[0000-0002-7188-1648]{Eric T. Wolf}
\affiliation{Laboratory for Atmospheric and Space Physics, Department
  of Atmospheric and Oceanic Sciences, University of Colorado,
  Boulder, CO, USA}

\author[0000-0001-7968-0309]{Colby Ostberg}
\affiliation{Department of Earth and Planetary Sciences, University of
  California, Riverside, CA 92521, USA}

\author[0000-0002-0139-4756]{Michelle L. Hill}
\affiliation{Department of Earth and Planetary Sciences, University of
  California, Riverside, CA 92521, USA}


\begin{abstract}

The discovery of terrestrial exoplanets is uncovering increasingly
diverse architectures. Of particular interest are those systems that
contain exoplanets at a variety of star--planet separations, allowing
direct comparison of exoplanet evolution (comparative
planetology). The Kepler-1649 system contains two terrestrial planets
similar both in size and insolation flux to Venus and Earth, although
their eccentricities remain largely unconstrained. Here we present
results of dynamical studies of the system and the potential effects
on climate. The eccentricities of the Kepler-1649 system are poorly
constrained, and we show that there are dynamically viable regions for
further terrestrial planets in between the two known planets for a
limited range of eccentricities. We investigate the effect of
eccentricity of the outer planet on the dynamics of both planets and
show that this results in high-frequency (1000--3000~years)
eccentricity oscillations in long-term stable configurations. We
calculate the resulting effect of these eccentricity variations on
insolation flux and present the results of 3D climate simulations for
the habitable zone planet. Our simulations demonstrate that, despite
large eccentricity variations, the planet can maintain stable climates
with relatively small temperature variations on the substellar
hemisphere for a variety of initial climate configurations. Such
systems thus provide key opportunities to explore alternative
Venus/Earth climate evolution scenarios.

\end{abstract}

\keywords{astrobiology -- planetary systems -- planets and satellites:
  dynamical evolution and stability -- stars: individual
  (Kepler-1649)}


\section{Introduction}
\label{intro}

Over the past several decades, thousands of exoplanets have been
discovered enabling statistical studies of exoplanet occurrence rates
\citep{catanzarite2011b,dressing2015b,kopparapu2013b,winn2015}. These
studies include the investigation of planets that lie within the Venus
Zone (VZ) \citep{kane2014e} and Habitable Zone (HZ)
\citep{kasting1993a,kane2012a,kopparapu2013b,kopparapu2014,kane2016c}. A
primary purpose of such studies is to place our solar system within
the context of planetary system architectures. One aspect of this
context is determining the frequency of systems that have terrestrial
planets in both the VZ and HZ. Such systems may provide potential
analogs for Venus and Earth, that in turn may provide key insights
into the divergence of the Venus/Earth atmospheric evolutions
\citep{kane2019d,ostberg2019}. Examples of such systems include
TRAPPIST-1 \citep{gillon2016,gillon2017a,luger2017b}, LHS~1140
\citep{dittmann2017b,ment2019}, and Kepler-186
\citep{quintana2014a}. Amongst the factors that influence the climate
evolution of terrestrial exoplanets are the eccentricity variations
\citep{barnes2013a,way2017a}, which may have contributed to the water
loss of Venus \citep{kane2020e}. Thus, a pathway to understanding the
major factors behind the Venus/Earth divergence is exploring the
dynamical impacts of eccentricity variations that may have occurred
(and continue to occur) in exoplanetary systems.

A particularly strong contributor to the currently known inventory of
terrestrial exoplanets was the {\it Kepler} mission
\citep{borucki2010a}, from which a catalog of HZ planets was
constructed \citep{kane2016c}. The Kepler-1649 system was discovered
by \citet{angelo2017a} which identified the observed planet as a
potential Venus analog due to the planetary size and similarity of
insolation flux to that of Venus. The prospect of the planet as a
Venus analog was strengthened by the detailed climate simulations of
\citet{kane2018d}, that showed the rapid rise in surface temperature
from a variety of initial conditions. The system became a strong
comparative planetology target when a further terrestrial planet was
discovered in the HZ by \citet{vanderburg2020a}. Thus, Kepler-1649 has
joined a select group of systems that are known to harbor terrestrial
planets in both the VZ and HZ. This class of system also provides
opportunities to study the role of complex dynamical evolution on the
orbital variation of planets \citep{smith2009,petrovich2015c},
including those within the HZ
\citep{kopparapu2010,kane2015b,kane2020b}.

In this paper we present a dynamical and climate study of the
Kepler-1649 system, describing the potential eccentricity variations
that result from the interactions between the planets, and the
possible impact on the climate of the HZ planet. Section~\ref{system}
describes the architecture of the system, provides estimates for the
planetary masses, and places constraints on the presence of additional
planets. Section~\ref{ecc} presents the results of dynamical
simulations that explore the amplitude and frequency of eccentricity
cycles through angular momentum exchanges between the planets. The
consequences of these dynamical interactions are discussed in
Section~\ref{climate}, including climate models of the surface
temperatures for various eccentricities and ocean models. These
climate simulations are provided for distinct eccentricities for
Kepler-1649c, which represent relatively short-lived climate states
for high-frequency eccentricity oscillations. We provide a discussion
of the results, their implications, and concluding remarks in
Section~\ref{conclusions}.


\section{System Architecture}
\label{system}

\begin{figure}
  \includegraphics[angle=270,width=8.5cm]{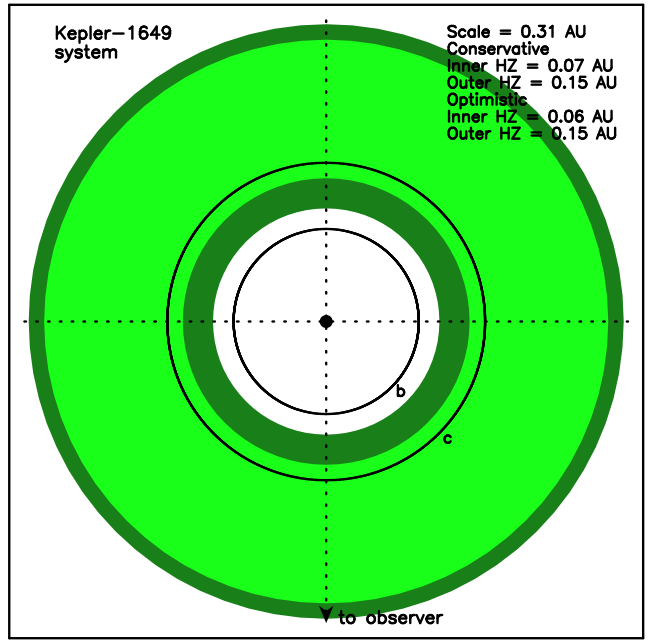}
  \caption{A top-down view of the Kepler-1649 system, where the star
    location is at the center of the crosshairs and the orbits of the
    known planets are indicated by solid circles. The extent of the
    CHZ is shown as the light green region and the OHZ extension to
    the CHZ is shown as the dark green region.}
  \label{hz}
\end{figure}

As described in Section~\ref{intro}, the Kepler-1649 system is
particularly fascinating from a comparative planetology aspect, since
it contains both potential analogs to Venus and Earth within the same
system. The stellar properties provided by \citet{vanderburg2020a}
include the mass ($M_\star = 0.1977 \pm 0.0051$~$M_\odot$), radius
($R_\star = 0.2317 \pm 0.0049$~$R_\odot$), luminosity ($L_\star =
0.00516 \pm 0.00020$~$L_\odot$), and effective temperature
$T_\mathrm{eff} = 3240 \pm 61$~K). The radii of the known planets in
the system are $R_p = 1.017 \pm 0.051$~$R_\oplus$ and $R_p =
1.06^{+0.15}_{-0.10}$~$R_\oplus$ for planets b and c respectively. To
calculate the semi-major axes of the planets, $a$, we use the
$a/R_\star$ information provided by \citet{vanderburg2020a}, resulting
in semi-major axes of $0.0482\pm0.0010$~AU and $0.0827\pm0.0017$~AU
for planets b and c respectively. The conservative HZ (CHZ) boundaries
are defined by the runaway green house limit at the inner edge and the
maximum greenhouse limit at the outer edge. The optimistic HZ (OHZ)
boundaries are determined empirically based on the assumption that
Venus may have had surface liquid water as recently as 1~Gya, and that
Mars may have surface liquid water $\sim$3.8~Gya. These boundaries are
described in more detail by \citet{kane2016c,kane2018a}. Based on the
above system properties, we calculated the extent of the CHZ and OHZ
boundaries for Kepler-1649 as 0.075--0.147~AU and 0.059--0.155~AU
respectively. The outer edge of the VZ corresponds to the inner edge
of the CHZ (the runaway greenhouse limit) at 0.075~AU. The extent of
the HZ and the orbits of the planets are represented in
Figure~\ref{hz}.

Although the orbits of the planets are shown as circular in
Figure~\ref{hz}, \citet{vanderburg2020a} do not constrain the
eccentricities in any significant way, though smaller planets tend to
have small eccentricities \citep{kane2012d}. \citet{vanderburg2020a}
further suggested that there could be an additional planet in an orbit
that lies between the two known planets. This suggestion was partially
motivated within the context of mean motion resonances (MMR) that may
exist within the system. The known b and c planets are close to the
9:4 MMR, which is a relatively weak resonance in terms of planetary
perturbations. However, an additional planet between the known planets
at a location of $\sim$0.063~AU would create a 3:2 resonant chain with
high stability potential \citep{batygin2013c}. Furthermore, an
analysis of Kepler systems by \citet{weiss2018a} found that planetary
architectures may preferentially contain similar sized planets
\citep{millholland2017c}, with the vast majority of Kepler planets
having a spacing greater than 10 mutual Hill radii. Assuming an Earth
mass for all three planets, a planet located at the 3:2 resonant chain
location between planets b and c would be $\sim$12.5 mutual Hill radii
away from either planet, consistent with the observed spacings in
other Kepler systems.

To investigate the prospect of an additional planet as a function of
eccentricity, we conducted a dynamical simulation using the {\sc
  REBOUND} orbital dynamics package \citep{rein2012a}. Such an
investigation requires an estimate of the planetary masses. There are
numerous planetary mass-radius relationships that have been developed
from which such an estimate may be derived
\citep[e.g.,][]{kanodia2019,zeng2019,neil2020,otegi2020,turbet2020b}. For
our dynamical simulations, we adopt the mass estimates provided by the
{\sc Forecaster} tool \citep{chen2017}, which produces mass
estimates of 1.06~$M_\oplus$ and 1.21~$M_\oplus$ for planets b and c
respectively.

\begin{figure*}
  \begin{center}
    \begin{tabular}{cc}
      \includegraphics[clip,width=8.4cm]{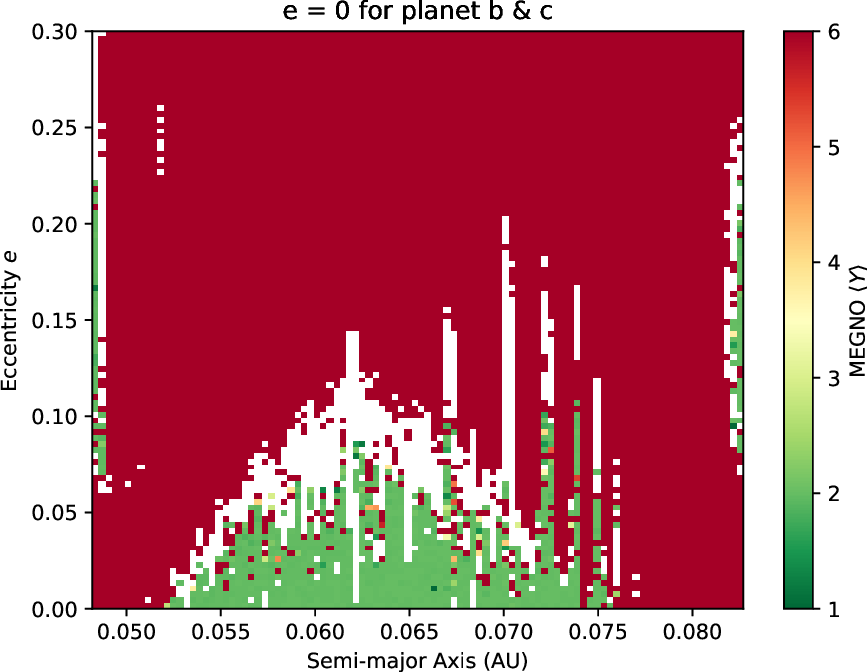} &
      \includegraphics[clip,width=8.4cm]{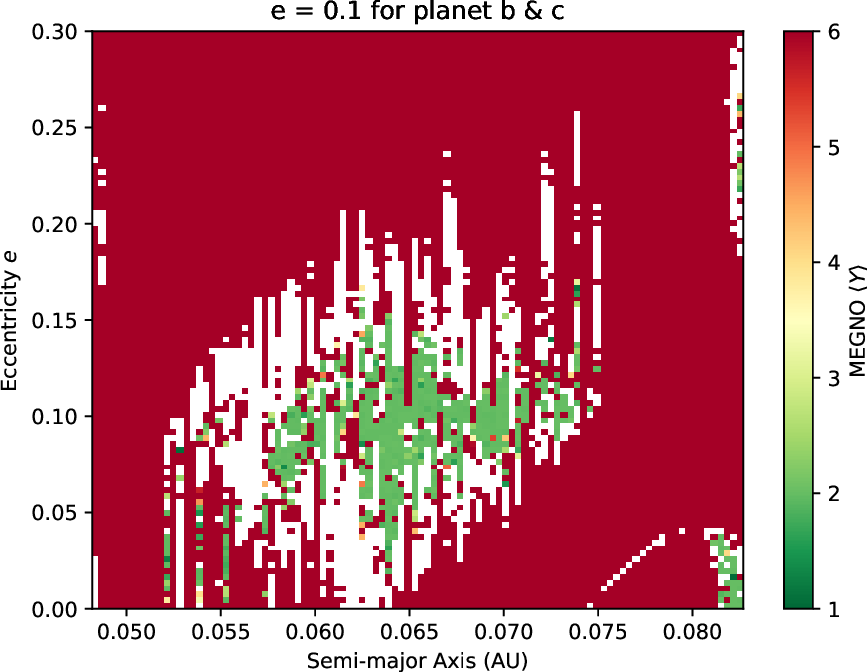} \\
      \includegraphics[clip,width=8.4cm]{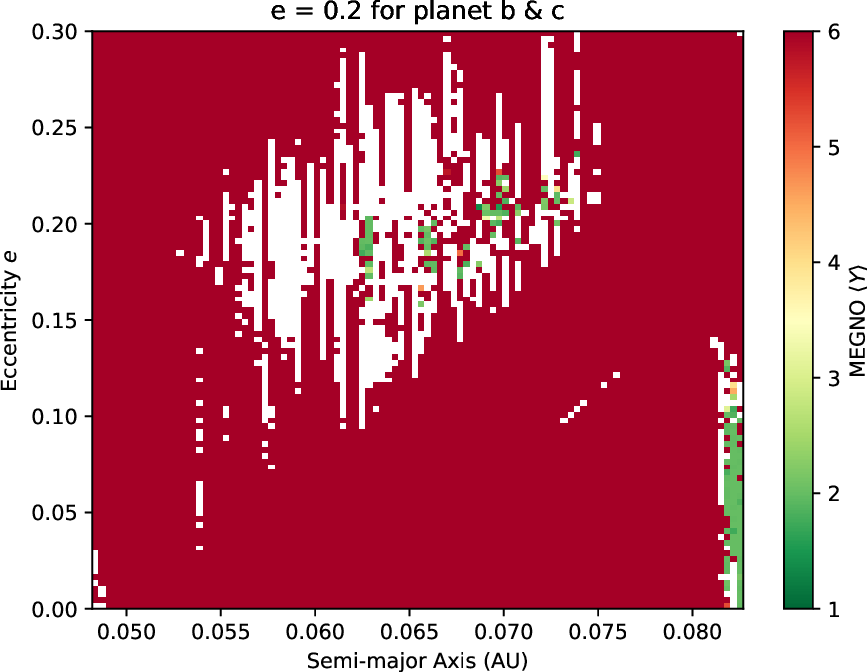} &
      \includegraphics[clip,width=8.4cm]{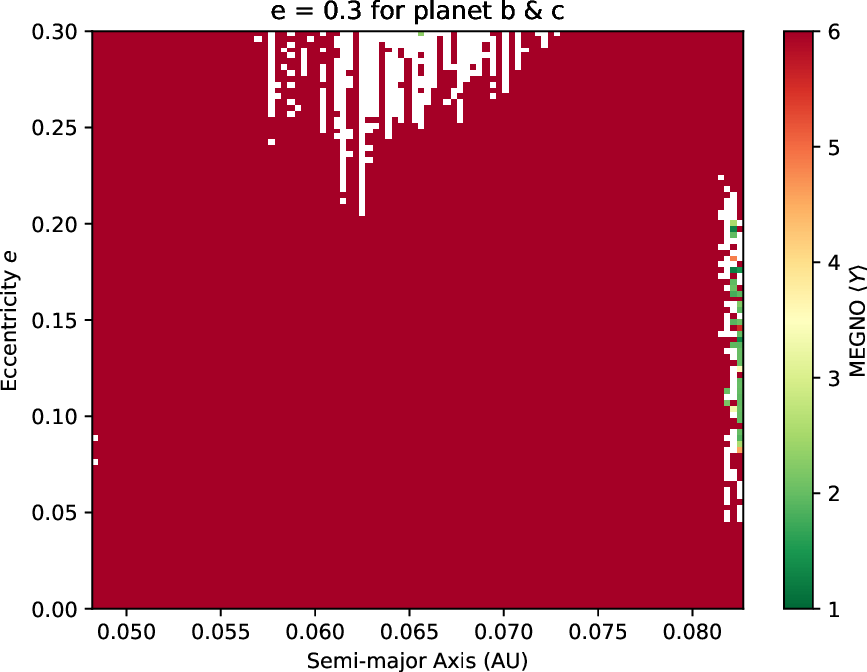} \\
      \includegraphics[clip,width=8.4cm]{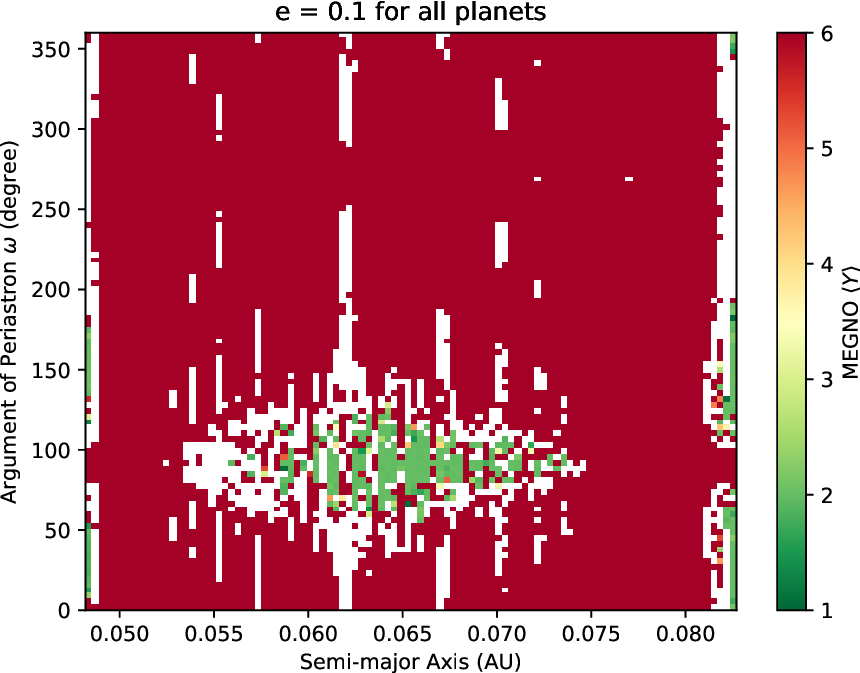} &
      \includegraphics[clip,width=8.4cm]{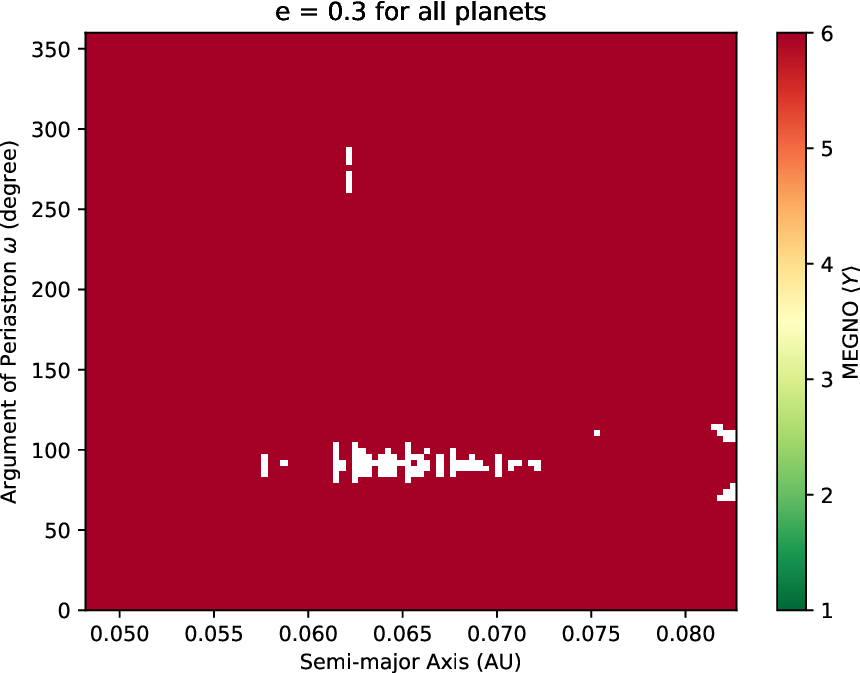} \\
    \end{tabular}
  \end{center}
  \caption{Results of a MEGNO analysis of the system, exploring the
    viability of an additional terrestrial planet in between planets b
    and c. This analysis was performed assuming initial eccentricities
    for planets b and c of 0.0 (top-left), 0.1 (top-right), 0.2
    (middle-left), and 0.3 (middle-right). The color scales indicate
    the final MEGNO value from the simulations, where green represents
    a stable system configuration, red represents an unstable
    (chaotic) system, and white represents an early termination of the
    simulation. Shown in the bottom panels are simulations that vary
    the argument of periastron for the injected planet for
    eccentricities of 0.1 (bottom-left) and 0.3 (bottom-right).}
  \label{megno}
\end{figure*}

The dynamical simulation was carried out using the Mean Exponential
Growth of Nearby Orbits (MEGNO) package within {\sc REBOUND} with the
symplectic integrator WHFast \citep{rein2015c}. MEGNO is a chaos
indicator that is useful in dynamical simulations to distinguish the
quasi-periodic or chaotic orbital time evolution of planetary
systems. The final numerical value returned by a MEGNO simulation
indicates whether the system would end up in a chaotic state or not,
where a chaotic state is less likely to maintain long-term
stability. For our simulation, we explored the possibility of an
additional planet in between the orbits of planet b and c by testing
how varying eccentricity and semi-major axis values for an injected
planet of one Earth mass affects the stochasticity of the system. We
conducted four sets of simulations that varied eccentricity, each
providing a grid of orbital parameters for the injected planet's
eccentricity (0.0--0.3) and semi-major axis (0.0482--0.0827~AU). Since
the eccentricities of the two known planets are unconstrained, we
tested planet b and c eccentricities of 0.0, 0.1, 0.2, and 0.3. Each
simulation duration was 10 million orbits for the outer most planet
with a time step of 0.001~years (0.365~days). The integration was set
to stop and return a large MEGNO value if the planet was ejected
beyond 100~AU. The results of our simulations for all four
eccentricity cases are shown in the top four panels of
Figure~\ref{megno}. The axes indicate the semi-major axes and
eccentricities tested for the injected planet in each of the four
cases. The colors correspond to the final MEGNO value for each
specific combination, with a MEGNO value $\sim$2 (green) indicating a
non-chaotic system \citep{hinse2010} and quasi-periodic motion for all
planets. A high MEGNO result, shown as red, indicates high chaos, and
irregular events that cause an early termination of the simulation,
such as close encounters and collision events, are shown in white. The
four simulations described above assumed an argument of periastron for
all three planets of 90$\degr$. We conducted further simulations that
varied the argument of periastron of the injected planet with respect
to those of the known planets. The results of these simulations for
the eccentricity 0.1 and 0.3 cases (for all planets) are shown in the
bottom two panels of Figure~\ref{megno}.

As can be seen in the top four panels of Figure~\ref{megno}, there is
a large parameter space between approximately 0.052~AU and 0.075~AU
with low eccentricity values where the injected Earth-mass planet
could be located with long-term stability. The largest range of
eccentricities for the injected planet that allow a stable
configuration occur at 0.063~AU which, as stated above, is the
location of the 3:2 MMR chain. The additional spikes visible in the
stability distribution, most obvious in the top-left panel, also
correspond to MMR locations, such as the 5:4 resonance location at
$\sim$0.071~AU. However, as the eccentricities for both planet b and c
increase to higher values, the stable locations for the injected
planet indicated by the green color gradually diminishes and the
locations shift to higher eccentricity values. When the eccentricities
for both planet b and c exceed 0.3, the stable locations for the
injected planet vanish, the reasons for which are discussed in
Section~\ref{ecc}. The bottom two panels of Figure~\ref{megno} show
that, even for the eccentricity case of 0.1 (bottom-left), the system
is most stable when the pericenters of the planets are approximately
aligned. As seen in the bottom-right panel of Figure~\ref{megno}, the
scarce regions of stability for the eccentricity case of 0.3 matches
the relative instability seen in the middle-right panel. It is worth
noting that the described simulations assumed system coplanarity, and
the introduction of mutual inclinations may present additional regions
of stability, particularly for scenarios that mitigate potential
overlap of orbits.


\section{Eccentricity Induced Dynamical Cycles}
\label{ecc}

As demonstrated in Section~\ref{system}, a non-zero eccentricity for
the known planets can place significant restrictions on the presence
of an additional terrestrial planet between them. Here we investigate
the time-dependent dynamical effects of non-zero eccentricities. Such
investigations have been carried out in numerous other works in a more
general context \citep{gladman1993,chambers1996,hadden2018b}. Our
N-body integrations were performed with the Mercury Integrator Package
\citep{chambers1999}, adopting a similar methodology to that described
by \citet{kane2014b,kane2016d,kane2019c}. The simulations used a
hybrid symplectic/Bulirsch-Stoer integrator with a Jacobi coordinate
system to provide increased accuracy for multi-planet systems
\citep{wisdom1991,wisdom2006b}. The time resolution of the simulations
was set to 0.2~days in order to comply with the recommendations of
\citet{duncan1998b} that require resolutions be $1/20$ of the shortest
orbital period (8.69~days for planet b).

We performed a series of dynamical simulations using the above
criteria that explore a range of eccentricities. Specifically, we
started each simulation with a circular orbit for planet b and
eccentricities for planet c that ranged from 0.0 to 0.5 in steps of
0.05. Each simulation was allowed to run for $10^6$ years, or
$\sim$$1.8\times10^7$ orbits of planet c, with orbital parameters for
both planets output every 10 simulation years. The suite of
simulations that explored the full range of starting eccentricities
for planet c were conducted using Earth masses for both planets b and
c, and also using the {\sc Forecaster} masses described in
Section~\ref{system}.

\begin{figure}
  \includegraphics[angle=270,width=8.5cm]{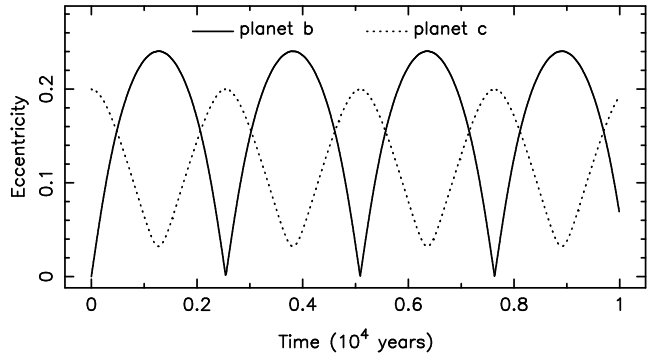}
  \caption{Eccentricity as a function of time for planet b (solid
    line) and planet c (dotted line) for the case of {\sc Forecaster}
    planet masses (1.06~$M_\oplus$ and 1.21~$M_\oplus$ respectively)
    and a starting eccentricity for planet c of 0.2.}
  \label{ecc1}
\end{figure}

\begin{figure*}
  \begin{center}
    \includegraphics[angle=270,width=16.0cm]{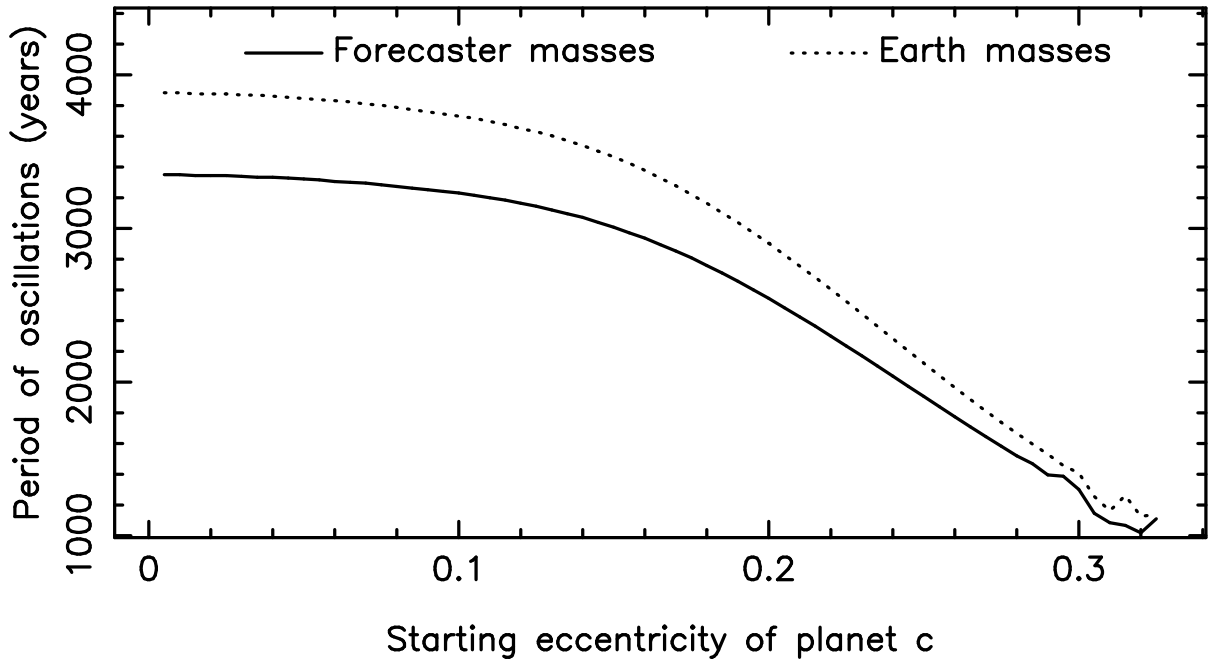}
  \end{center}
  \caption{The period of the eccentricity oscillations of planets b
    and c as a function of the starting eccentricity for planet c, for
    both the {\sc Forecaster}-mass (solid line) and Earth-mass (dotted
    line) cases. Note that the system is dynamically unstable for
    eccentricities beyond $\sim$0.3.}
  \label{ecc2}
\end{figure*}

The full set of dynamical simulations demonstrated that the system is
able to remain stable for $10^6$ years up to a starting eccentricity
for planet c of 0.325, beyond which the system rapidly loses dynamical
integrity. The 0.325 eccentricity upper limit was true for both the
Earth-mass and {\sc Forecaster}-mass scenarios. In particular, we
found that the eccentricity of planet c induces a near-equivalent
eccentricity in planet b, whereby each planet oscillates in
eccentricity in order to conserve the system angular momentum. For
example, shown in Figure~\ref{ecc1} are the time-dependent
eccentricities of planet b (solid line) and planet c (dotted line)
where the assumed masses are those from {\sc Forecaster} and the
starting eccentricity of planet c is 0.2. The figure demonstrates the
full amplitude and frequency of the eccentricity variations over a
relatively brief period of $10^4$ years. The period of the
eccentricity oscillations in this case is $\sim$2500~years.

To investigate this further, we conducted a fourier analysis of each
of the dynamical simulation outputs described above. The periods of
the eccentricity oscillations, for both {\sc Forecaster} and Earth
masses, as a function of starting planet c eccentricity are shown in
Figure~\ref{ecc2}. There are several aspects to note about this
figure. Firstly, the period of the oscillations decreases with
increasing eccentricity. Secondly, the period of the oscillations is
systematically smaller for the larger {\sc Forecaster}
masses. Thirdly, the dynamical instability that results beyond
eccentricities of 0.3 is clearly visible and is likely related to the
diminishing oscillation period with increasing eccentricity. These
simulation results agree well with similar calculations that may be
performed using the analytical expressions derived by
\citet{hadden2018b}. It is worth noting that the period of the
eccentricity oscillations and the eccentricity at which instability
occurs is likely influenced by the proximity of the planets to the 9:4
MMR \citep{vinson2018,hadden2019b}, discussed in
Section~\ref{system}. Additional factors include the mutual
inclinations of the planets combined with non-zero stellar obliquity
\citep{spalding2018b}. Regardless, the high frequency of eccentricity
oscillations may result in significant climate consequences for the
planets.


\section{Dynamical Consequences on Climate}
\label{climate}

\begin{figure}
  \includegraphics[angle=270,width=8.5cm]{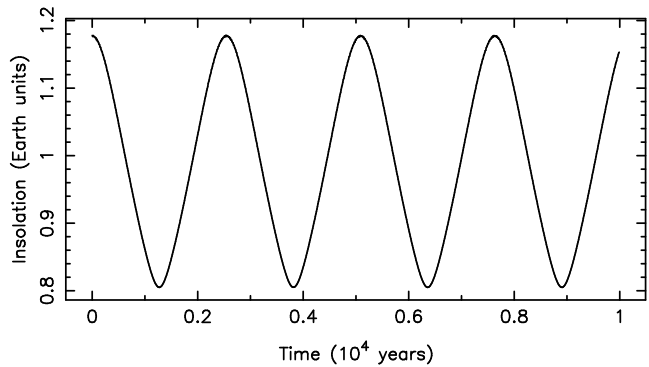}
  \caption{Maximum flux (insolation) received by planet c during a
    single orbit (in Earth units) as a function of time for an
    eccentricity of 0.2.}
  \label{fluxfig}
\end{figure}

Here we discuss the implications of the variable eccentricity for the
possible climate scenarios of planet c. Given the system parameters
provided in Section~\ref{system}, the insolation flux received by
planet c at its semi-major axis is $\sim$75\% of the mean Earth
insolation flux. The eccentricity oscillations described in
Section~\ref{ecc} will result in a correspondingly variable insolation
flux for the planet, whose amplitude and period depend on the
eccentricity. For example, consider the case of a starting
eccentricity of 0.2 for planet c, as shown in Figure~\ref{ecc1}. The
variation in the maximum insolation flux of the planet with time is
shown in Figure~\ref{fluxfig}. Since, as shown in Figure~\ref{ecc1},
the eccentricity of planet c never reaches zero, the maximum
insolation flux is always larger than that for a circular
orbit. Figure~\ref{fluxfig} shows that the maximum insolation in this
case oscillates approximately around an Earth equivalent flux with a
variation of $\pm 20$\%.

The Resolving Orbital and Climate Keys of Earth and Extraterrestrial
Environments with Dynamics ({\sc ROCKE-3D}), described in detail by
\citet{way2017b}, is a three-dimensional climate system model
developed at the NASA Goddard Institute for Space Studies (GISS)
specifically for studying planetary and exoplanetary atmospheres. We
use {\sc ROCKE-3D} to constrain the climatic effects of oscillations
in the orbital-rotational properties of Kepler-1649c. {\sc ROCKE-3D}
has previously been used for studying the climate and habitability of
terrestrial extrasolar planets, including studies of Proxima Centauri
b \citep{delgenio2019a}, moist and runaway greenhouse states
\citep{fujii2017b,way2018,kane2018d}, the effects of planet
orbital-rotational properties \citep{way2017a,colose2019}, and the
effects of ocean circulation
\citep{checlair2019a,olson2020,salazar2020}. While the atmospheric
composition and surface properties of Kepler-1649c remain
unconstrained, here we have conducted a set of simulations using {\sc
  ROCKE-3D} to illuminate the possible role of eccentricity evolution
on climate. Following from Figure~\ref{ecc2}, for a starting
eccentricity of 0.3, Kepler-1649c could experience eccentricity
oscillations on $\sim$1200 year timescales. To constrain the bounds of
the potential climate states experienced by Kepler-1649c during its
dynamical evolution, we simulate scenarios with zero eccentricity and
synchronous rotation, and with an eccentricity of 0.3 and a 3:2
resonant rotation state. Note, for these cases we are not simulating
the continuous evolution of Kepler-1649c over many thousands of years,
but rather we are simulating the equilibrium climate states for
potential end-member orbital-rotational configurations.  However, if
the spin dynamics become chaotic, these end-member rotation states may
not be fully realized \citep{correia2004}. Thus, we also conduct
simulations that assume a constant, non-resonant asynchronous spin
rate in order to approximate chaotic spin dynamics. For these cases,
we assume a 16 Earth-day rotation period, which is an intermediate
value between the synchronous rotation period of $\sim$19.535 days,
and the 3:2 resonant rotation period of $\sim$13.0235 days. Each of
our simulations achieve radiative equilibrium after 100--200 years,
well within the timescales of possible eccentricity variations
described in Section~\ref{ecc}. This is also comparable to the
timescales provided by the continuously variable eccentricity approach
adopted by \citet{way2017a}.

As described in Section~\ref{system}, Kepler-1649c lies within the
CHZ, close to the inner edge. In all climate simulations conducted
here we assume a globally ocean covered planet with no emergent
continents. While this represents only one of many potential surface
configurations for Kepler-1649c, the study of ocean covered worlds
provides a first-order approximation for the study of climate
evolution on potentially habitable worlds. We test two different
assumptions for the ocean layer. "Slab" ocean simulations assume a
100~m deep ocean layer with no horizontal transports, but with the
thermodynamic exchange of heat and moisture between the ocean and
atmosphere. "Dynamic" ocean simulations utilize a 900~m deep fully
coupled ocean circulation model capable of predicting horizontal and
vertical transports, in addition to thermodynamic exchanges with the
atmosphere. Both ocean configurations allow for sea-ice
formation. Salinity is held equal to that of the modern Earth's ocean.
Atmospheric compositions also remain unconstrained from observations
and therefore constitute a large uncertainty on the climate
state. Here, we choose two nominal atmospheric compositions which are
"Earth-like" (1~bar total with 400 ppm CO$_2$ and an N$_2$
background), and a 1~bar CO$_2$ atmosphere. For all climate
simulations, water vapor and clouds are variable constituents in the
atmosphere, controlled by temperature, evaporation, convection,
advection, and condensation. Our selections for atmospheric
compositions were taken from the test cases for habitable exoplanets
proposed by \cite{fauchez2020b}, that facilitate the study of a
diversity of climate states. For planets residing in the middle of the
habitable zone, as does Kepler-1649c, an Earth-like atmospheric
composition will result in a cold climate state, dominated by ice and
where ocean transport processes are of particular importance
\citep{delgenio2019a}, while a 1~bar CO$_2$ atmosphere will result in
a warm climate state, dominated by the hydrological cycle, clouds, and
atmospheric transport \citep{yang2019a}.  Thus the selection of these
two atmospheric compositions allows us to capture some diversity of
habitable zone climates. A wide ranging parameter study of the
infinite variety of possible surface and atmospheric compositions for
Kepler-1649c is beyond the scope of this work. In total, we have
conducted 16 different climate simulations of Kepler-1649c, with
results summarized in Table~\ref{tab:climate_table}.

\begin{figure*}
  \begin{center}
    \includegraphics[width=16cm]{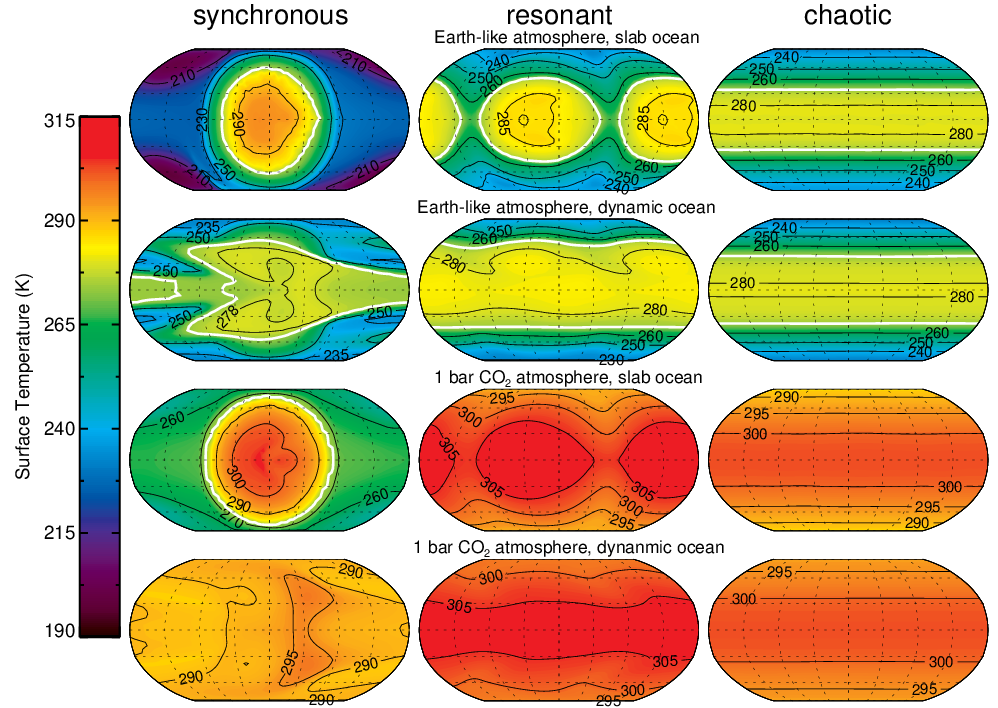}
  \end{center}
  \caption{Surface temperatures maps from our climate simulations of
    Kepler-1649c. The left column assumes synchronous rotation with
    zero eccentricity, the center column assumed 3:2 resonant rotation
    with an eccentricity of 0.3, and the right column assumes chaotic
    rotation approximated by a 16 day asynchronous orbit.  For the
    chaotic states, eccentricity does not have a significant effect on
    the climate and thus zero and 0.3 eccentricity results are
    averaged together. Rows (labeled) illustrate the different surface
    and atmospheric configurations. All simulations were performed with
    {\sc ROCKE-3D}.}
  \label{climatefig}
\end{figure*}

Figure~\ref{climatefig} shows surface temperature maps from our
climate simulations with {\sc ROCKE-3D}. The results shown are from an
average of the last 100 orbits of Kepler-1649c (i.e. 1950 Earth days,
5.3 Earth years). The bold white line indicates where the surface
temperature is 273~K, and thus serves as a close proxy for the sea ice
margin. The columns show climate results from different rotation
states, including synchronous, resonant, and chaotic. Synchronous
rotation assumes zero eccentricity and a rotation period equaling the
orbital period. Resonant rotation assumes a rotation period in a 3:2
resonance with the orbital period and an eccentricity of 0.3. Chaotic
rotation, as mentioned previously, is approximated by a 16 day
asynchronous rotation period. The global mean surface temperature and
surface temperature maps are largely unchanged for chaotic rotators
with eccentricities of zero or 0.3. Thus, plotted are an average
between these two states. Finally, each row in figure~\ref{climatefig}
illustrates different atmosphere and surface combinations, including
Earth-like and 1 bar CO$_2$ atmospheres, with slab and dynamic ocean
models.

\begin{table*}
\centering
\begin{tabular} {|c|c|c|c|c|c|c|}
\hline
\# & Rotation State & Atmosphere & Ocean & T$_S$ mean (K) &  Ocean (\%) & Albedo \\
\hline
1  & 1:1,    $e=0.0$ & Earth-like & slab & 244.7 & 25.1 & 0.303 \\
\hline
2  & 3:2,    $e=0.3$ & Earth-like & slab & 268.0 & 57.9 & 0.223 \\
\hline
3  & 16 day, $e=0.0$ & Earth-like & slab & 266.0 & 61.2 & 0.199 \\
\hline
4  & 16 day, $e=0.3$ & Earth-like & slab & 268.9 & 70.1 & 0.215 \\
\hline
\hline
5  & 1:1,    $e=0.0$ & Earth-like & dynamic & 263.4 & 62.5 & 0.244 \\
\hline
6  & 3:2,    $e=0.3$ & Earth-like & dynamic & 272.3 & 79.2 & 0.210 \\
\hline
7  & 16 day, $e=0.0$ & Earth-like & dynamic & 267.6 & 68.6 & 0.187 \\
\hline
8  & 16 day, $e=0.3$ & Earth-like & dynamic & 269.6 & 73.0 & 0.210 \\
\hline
\hline
9  & 1:1,    $e=0.0$ & 1 bar CO$_2$ & slab & 275.7 & 39.2 & 0.239 \\
\hline
10 & 3:2,    $e=0.3$ & 1 bar CO$_2$ & slab & 304.1 & 100 & 0.245 \\
\hline
11 & 16 day, $e=0.0$ & 1 bar CO$_2$ & slab & 296.8 & 100.0 & 0.258 \\
\hline
12 & 16 day, $e=0.3$ & 1 bar CO$_2$ & slab & 301.2 & 100.0 & 0.272 \\
\hline
\hline
13 & 1:1,    $e=0.0$ & 1 bar CO$_2$ & dynamic & 291.3 & 100 & 0.269 \\
\hline
14 & 3:2,    $e=0.3$ & 1 bar CO$_2$ & dynamic & 304.8 & 100 & 0.246 \\
\hline
15 & 16 day, $e=0.0$ & 1 bar CO$_2$ & dynamic & 297.0 & 100.0 & 0.262 \\
\hline
16 & 16 day, $e=0.3$ & 1 bar CO$_2$ & dynamic & 303.4 & 100.0 & 0.262 \\
\hline
\end{tabular}
\caption{Basic climatic statistics from 3D climate simulations of
  Kepler-1649c with {\sc ROCKE-3D}.}
\label{tab:climate_table}
\end{table*}

Global mean surface temperatures range between 244.7~K and 304.8~K,
and surface liquid water is maintained over at least some of the
planet surface in all of the cases simulated
(Table~\ref{tab:climate_table}). Thus, all climates simulated are at
least partially habitable, with many states being globally
habitable. If the orbital-rotational state of Kepler-1649c can fully
oscillate between synchronous and resonant modes, significant temporal
changes to the global mean surface temperatures and the total
habitable area of the planet, which here we use the open ocean
fraction as a proxy for the habitable surface area, would occur. The
shift from synchronous rotation to resonant rotation affects climate
both by changing the longitudinal patterns of the stellar flux
incident on the planet, and also by changing the Coriolis parameter
which feeds back on atmospheric and ocean circulations. The breaking
of synchronicity and the increase in the rotation rate tends to result
in increased zonal transports resulting in longitudinally banded
patterns, as opposed to the classic eyeball climate states found for
slow and synchronous rotators
\citep{pierrehumbert2011a,turbet2016,kopparapu2017,haqqmisra2018a}. The
inclusion of non-zero eccentricity with a 3:2 orbital resonance
produces a characteristic double hot-spot evident in temporal mean
climate statistics, due to the fact that two opposite faces of the
planet are periodically irradiated at perihelion
\citep{boutle2017,delgenio2019a}. The double hot spot pattern is
clearly evident in the slab ocean cases. The orbital period of
Kepler-1649c is short (19.5 days) compared to thermal timescales of
the ocean, thus opposing sides of the planet remain permanently
warm. For dynamic ocean cases, a faint signal of the double hot spot
phenomenon persists, but zonal heat transports by the ocean results in
nearly zonally uniform patterns of surface temperature
(Figure~\ref{climatefig}).

For corresponding end-member scenarios of synchronous and resonant
modes (Table~\ref{tab:climate_table}), global mean surface
temperatures could change by as little as 8.9~K for an Earth-like
atmosphere with dynamic ocean circulation or by as much as 28.4~K for
a CO$_2$ dominated planet with a slab ocean. For both atmospheric
compositions, the climate sensitivity between end-member
orbital-rotational states is reduced when considering a full dynamic
ocean compared to the slab ocean cases. Perhaps surprisingly, the
optically thicker 1 bar CO$_2$ simulations are more sensitive to the
orbital-rotational state than are the Earth-like atmosphere. While
further investigation is warranted, we surmise that this is due to
more vigorous water vapor and cloud feedbacks that occur for our 1 bar
CO$_2$ atmospheres, due to their generally warmer climate states
compared to the Earth-like atmospheres which remain cold and ice
dominated.

However, as noted above, these end-member scenarios may never be fully
realized. The relatively rapid changes in orbital-rotational states
may instead drive Kepler-1649c into chaotic rotation, which would
result in asynchronous planet rotation and not into any favored
resonant state. In such a scenario, the climate patterns of
Kepler-1649c become zonally uniform for all combinations of
atmospheric composition and ocean assumptions (Figure~\ref{climatefig}
third column), as starlight would irradiate all longitudes of the
planet equally with no preferred faces receiving more insolation on
time-average than any other.

In all cases simulated, the change in surface temperature caused by
the dynamical evolution is relatively small on the substellar
hemisphere, but is much larger for the antistellar hemisphere. Here,
we define substellar and antistellar hemispheres based on the
synchronous rotating orbital-rotational reference frame.  Naturally,
for non-synchronous cases (resonant and chaotic), all sides of the
planet receive periodic stellar flux, meaning that the opposing
hemisphere is no longer permanently cloaked in darkness and enjoys its
moment in the starlight, resulting in considerable warming. The
substellar-antistellar dichotomy is most strongly seen for the slab
ocean cases. In both synchronous and resonant rotating instances, the
inclusion of dynamic ocean circulations effectively transports heat
from the day-side to the night-side of the planet, muting the
substellar-antistellar dichotomy.

Our 3D climate simulations reveal that the modulation of
orbital-rotational states does not trigger true climate catastrophes
for Kepler-1649c (i.e., runaway glaciation or runaway greenhouse
states) given the atmospheric compositions explored here. If
Kepler-1649c oscillates between synchronous and resonant states,
significant millennial scale climate changes would occur that are far
in excess of anything that the Earth has experienced over such a short
timescale. Given the timescales of variation here (Figure~\ref{ecc1}),
life as we know it may struggle to adapt to such wide temperature
variations indicated by our simulations. Extant lifeforms may remain
confined to the substellar hemisphere due its relative climatic
stability compared to the significantly more intense variations that
occur on the antistellar hemisphere. However, if instead Kepler-1649c
settles into chaotic rotation, millennial scale climate changes would
be negligible.

Other factors not explored here may also affect the sensitivity of
Kepler-1649c's climate to oscillations in its orbital-rotational
state. Here we have modeled only selected states in equilibrium. If
the orbital-rotational state of Kepler-1649c instead evolves
monotonically between our illustrated end-member cases, the climate
states may instead gradually smear between synchronous and resonant.
Chaotic rotation states may also experience time evolution of the
rotation period. However, differences in non-resonant asynchronous
rotation periods are not likely to drive significant changes in
climate, compared to toggling into, and out of, synchronous and
resonant states.

The atmospheric composition also presents a significant unconstrained
parameter. While the atmospheric compositions studied here all result
in temperate climates, if the atmosphere of Kepler-1649c were
considerably different, the climate sensitivity between states may
deviate considerably. For instance, if Kepler-1649c possesses an
overwhelmingly dense, cloud and haze filled atmosphere, the surface
may receive little direct starlight and become decoupled from changes
in the patterns of stellar flux, and surface temperatures may trend
toward a globally uniform state regardless of rotation state
\citep{turbet2016}. Alternatively, if Kepler-1649c instead has a dry,
clear, optically transparent atmosphere, the change in rotation state
may also cause little difference to the surface temperature as the
planet's energy balance and thus climate would be governed simply by
the albedo and emissivity properties of the planet's solid surface,
along with the instantaneous stellar flux input. Finally, our results
indicate that the ocean treatment makes a considerable difference in
the modulation of surface temperatures between synchronous and
resonant states, but makes little difference for the chaotic rotation
state. Still, changes in the ocean depth, and the presence and
location of continents would significantly influence the results shown
here through the modulation of ocean circulations
\citep{yang2019a}. Note, \cite{leconte2018b} argued that tidally
locked planets with emergent continents will tend to have their
continents migrate to the substellar or antistellar points. If
continents cluster near the substellar point, recent work with {\sc
  ROCKE-3D} suggests that this would mute the effects of ocean
circulations \citep{salazar2020}. The varied location of continents
remains a significant uncertainty in modern terrestrial exoplanet
climate modeling, due to their tight coupling to ocean circulations,
ocean heat transport, and ultimately climate.


\section{Conclusions}
\label{conclusions}

A major focus of exoplanetary studies lies in the exploration of the
full diversity of planetary system architectures. However, some of
these architectures are similar enough to our own solar system such
that they may shed insight into the orbital and atmospheric evolution
of our terrestrial planets. Thus, systems that contain terrestrial
planets within both the VZ and the HZ are of particular interest for
the investigation of potential Venus/Earth analogs that exist within
similar yet different contexts of age, composition, host star type,
and orbital dynamics.

The Kepler-1649 system is one of the few known systems to harbor such
a potential Venus/Earth analog, where the similar age and, presumably,
composition of the planets allows the removal of several free
parameters from which other comparative studies may be conducted. Here
we have presented the results of dynamical analysis in which we have
demonstrated that additional terrestrial planets could exist in orbits
that lie between the two known planets. This is not an unlikely
scenario given the dynamical packing observed for compact planetary
systems \citep{fang2013,winn2015} and the necessity of coplanarity for
transits to be observed \citep{fang2012f,ballard2016}. Additionally,
the locations of significance resonance, such as the potential 3:2 MMR
chain described in Section~\ref{system}, lend credence to a possible
planet located between the known b and c planets. Our dynamical
simulations further demonstrate that varying eccentricity can be used
as a tool to exclude either eccentricities of the known planets or the
presence of the hypothetical additional planet. We also explored the
effect of eccentricity on the dynamical evolution of the two known
planets, the results of which show that the system is stable up to
eccentricities of $\sim$0.3 and that the frequency of the eccentricity
oscillations can be exceptionally high (1000--3000 years). Such
high-frequency eccentricity oscillations are a fundamental component
of Milankovitch cycles for exoplanets
\citep{deitrick2018a,deitrick2018b,horner2020a} and may play an
important role in climate evolution.

Our {\sc ROCKE-3D} simulations for the climate of Kepler-1649c under
various eccentricity and ocean circulation assumptions demonstrate
that in fact the surface temperature variation remains relatively
small, even for eccentricity at the high end. The region of highest
climate stability consistently lies in the substellar hemisphere for
all of our simulations, indicating a potential longitude dependence of
metabolic processes for any life that may exist on such a
planet. Thus, the overall combination of both dynamical and climate
evolution studies for this planetary system show that even though
their orbits may change far more rapidly than that observed within the
Solar System, atmospheric characterization may yet reveal robust
environments that maintain temperate surface conditions. Such
dynamically active systems should therefore not be ruled out in the
continuing search for potentially habitable worlds.


\section*{Acknowledgements}

The authors would like to thank Michael Way for valuable feedback on
the climate simulation methodology, and the anonymous referee for the
useful corrections. This research has made use of the Habitable Zone
Gallery at hzgallery.org. The results reported herein benefited from
collaborations and/or information exchange within NASA's Nexus for
Exoplanet System Science (NExSS) research coordination network
sponsored by NASA's Science Mission Directorate.


\software{Mercury \citep{chambers1999}, REBOUND \citep{rein2012a},
  ROCKE-3D \citep{way2017b}}




\end{document}